\documentstyle[prl,twocolumn,aps,epsf]{revtex}

\begin{document}
\twocolumn[\hsize\textwidth\columnwidth\hsize\csname @twocolumnfalse\endcsname
\draft
\tolerance 500
\title{Aging effects in the quantum dynamics of a dissipative free particle: non-ohmic case}
\author{Alain Mauger$^{1}$ and No\"elle Pottier$^{2}$}

\address{$^1$ Laboratoire des Milieux D\'esordonn\'es et H\'et\'erog\`enes,
CNRS UMR 7603, Universit\'{e} Paris 6,\\
4, place Jussieu, 75252 Paris Cedex 05, France}

\address{$^2$ Groupe de Physique des Solides, CNRS UMR 7588,
Universit\'{e}s Paris 6 et Paris 7,\\
2, place Jussieu, 75251 Paris Cedex 05, France}
\maketitle

\begin{abstract}
We report new results related to the two-time dynamics of the coordinate of a quantum free particle, damped through
its interaction with a fractal thermal bath (non-ohmic coupling $\sim\omega^\delta$ with $0<\delta<1$ or $1<\delta<2)$.
When the particle is localized, its position does not age. When it undergoes anomalous diffusion, only its displacement may be
defined. It is shown to be an aging variable. The finite temperature aging regime is self-similar. It is described by a
scaling function of the ratio ${t_w/\tau}$ of the waiting time to the observation time, as characterized by an exponent
directly linked to $\delta$. 

\end{abstract}

\pacs{PACS numbers: 05.30.-d, 05.40.+j, 02.50.Ey}

\vskip2pc]

In complex out-of-equilibrium systems such as structural glasses and spin glasses, the trend towards equilibrium is
accompanied by aging effects: for instance the two-time correlation functions of certain out-of-equilibrium dynamic variables
may not be invariant by time translation, even in the limit of a large waiting time (or age). The fluctuation-dissipation
theorem, which is valid for dynamic variables at equilibrum, is not verified in this case. Actually the study of aging effects
and of the related violation of the fluctuation-dissipation theorem (FDT) is a fundamental problem of the physics of
dissipative out-of-equilibrium systems${\protect\cite{Bouchaud}}$. Several aspects of this problem still stand as largely open
questions, especially in the case of quantum dissipative systems.
 
In order to be able to discuss these questions at any temperature, one has to work within a quantum framework, the time
scale ${\hbar/kT}$ playing a crucial role in the low-temperature dynamics. Since aging effects are also encountered, not
only in complex systems, but also in simpler systems, nor disordered nor frustrated$\protect\cite{Cugliandolo}$, it is very
natural, to begin with, to carry out the study of quantum aging in this latter type of systems.

One archetype of a simple quantum dissipative system displaying aging effects is a particle coupled to a thermal bath but
otherwise free, that is evolving in the absence of potential. Aging effects on the displacement correlation function and
the corresponding violation of the quantum FDT have recently been discussed$\protect\cite{Pottier}$ for a specific model of
dissipation, namely the so-called ohmic model. It corresponds to a particle undergoing standard quantum Brownian
motion${\protect\cite{Hakim,Aslangul1}}$.  The damped ohmic equation of motion can be given a non retarded form in the
classical limit, in which diffusion is normal. However, the ohmic model does not allow to handle all the physical situations
of interest, for instance those in which the damped motion is described by a truly retarded equation even in the classical
limit and in which either localization or anomalous diffusion phenomena are taking place. 

In this Letter, we examine aging effects in this simple system using for the dissipation a versatile model able to generate
various damped equations of motion, either instantaneous or retarded in the classical limit. The dissipation is introduced
{\sl via\/} a linear coupling of the particle to a thermal bath having a continuous distribution of modes of bandwidth
$\omega_c$ (Caldeira and Leggett model${\protect\cite{Caldeira,Leggett}}$). The central ingredient of the model is the product
of the density of modes of the bath $g(\omega)$ times the squared coupling constant
$|\lambda(\omega)|^2$, a product assumed to vary as $\omega^\delta$ at frequencies $\omega\ll\omega_c$. In the ohmic model, the
dissipative exponent $\delta$ is equal to 1. The algebraic cases $0<\delta<1$ and $\delta>1$ are known,
respectively, as the subohmic or superohmic dissipation models${\protect\cite{Schramm,Grabert,Aslangul2,Weiss}}$. 

In the following we first show that for $0<\delta<2$ the particle velocity equilibrates at large times and does not age. Then,
turning to the study of the particle coordinate, we discuss the domain of ($\delta,T$) parameters for which aging is
taking place, or not. Actually the question is a  non-trivial one. Indeed the two following properties have been
demonstrated$\protect{\cite{Schramm,Grabert,Aslangul2,Weiss}}$. First, at $T=0$ for $0<\delta<1$, the particle is localized,
in the sense that the mean square displacement $\Delta x^2(t)=\langle[x(t)-x(0)]^2\rangle$ tends towards a constant at
infinite time. Second, at $T=0$ for $1\leq\delta<2$, and also at finite $T$ for $0<\delta<2$, $\Delta x^2(t)$ diverges at
infinite time, the diffusion being anomalous except at finite $T$ for $\delta=1$. As a result, we find clear-cut behaviors: 
as far as the coordinate is concerned, the localized particle does not age, while the diffusing one ages. In this latter
situation we provide analytic expressions for the effective temperature and the associated fluctuation-dissipation ratio
allowing to write a modified FDT at finite temperature. We show in particular that the aging regime which emerges at large
times is self-similar.

In the Caldeira and Leggett model, the particle is coupled linearly to a set of harmonic oscillators in thermal equilibrium at
temperature $T$. The Hamiltonian of the particle-plus-bath system reads, in obvious notations,
%%%%%%%%%%%%%%%%%%%%%%%
\begin{eqnarray}
\nonumber H={p^2\over 2m}-x\sum_\nu\lambda_\nu(b_\nu&+&b_\nu^\dagger)+\\
&&\sum_\nu\hbar\omega_\nu
b_\nu^\dagger b_\nu+ x^2\sum_\nu{\lambda_\nu^2\over\hbar\omega_\nu},
\label{hamil}
\end{eqnarray}
%%%%%%%%%%%%%%%%%%%%%%%
where $\lambda_\nu$ is a real coupling constant. For $\omega>0$, the quantity $2\pi\,g(\omega)\,|\lambda(\omega)|^2={1\over
2}m\hbar\omega K(\omega)$ is modelized as
%%%%%%%%%%%%%%%%%%%%%%%
\begin{equation}
\label{modelization}
K(\omega)=2\gamma{\bigl({\omega\over\gamma}\bigr)}^{\delta-1}f_c\bigl({\omega\over\omega_c}\bigr),
\end{equation}
ä%%%%%%%%%%%%%%%%%%%%%%
where $\gamma$ is a coupling frequency and $f_c$ a high-frequency cut-off function of
typical width $\omega_c$. The definition of $K(\omega)$ is extended to $\omega<0$ by imposing that it must be an even function
of $\omega$. The particle position operator obeys the retarded equation of motion
%%%%%%%%%%%%%%%%%%%%%%%
\begin{equation}
\label{eqofmotion}
\ddot x(t)+\int_{t_i}^tdt'k(t-t')\dot x(t')=
-x(t_i)k(t-t_i)+{1\over m}F(t),
\end{equation}
%%%%%%%%%%%%%%%%%%%%%%%
where $t_i$ denotes the initial time at which the coupling is switched on. In Eq. (\ref{eqofmotion}), the inverse Fourier
transform~$k(t)$ of $K(\omega)$ plays the role of a memory kernel, and $F(t)$ is a linear combination of
bath operators, acting as a stationary random force of correlation function
%%%%%%%%%%%%%%%%%%%%%%%
\begin{equation}
C_{FF}(t)=m\int_{-\infty}^\infty{d\omega\over 2\pi}\,\Re {\rm e}\,\tilde K(\omega)\hbar\omega\coth{\beta\hbar\omega\over
2}e^{-i\omega t}
\end{equation}
%%%%%%%%%%%%%%%%%%%%%%%
with $\Re {\rm e}\,\tilde K(\omega)={1\over 2}K(\omega)$ and $\beta=(k_BT)^{-1}$.

It has been demonstrated that, for $0<\delta<2$ the total mass of the particle and of the bath oscillators
diverges, while for $\delta>2$ it remains finite and can be considered as a
renormalized mass$\protect{\cite{Schramm,Grabert,Aslangul2,Weiss}}$. For $0<\delta<2$ ($\delta\neq
1$),\break the average particle velocity $\langle v(t)\rangle$ relaxes towards zero at large $t-t_i$ like $(t-t_i)^{\delta-2}$
(the average is taken over the variables of the initial state of the particle and over the bath variables). For
$\delta=1$, the relaxation is exponential. The initial value $\langle v(t_i)\rangle$ being forgotten for $0<\delta<2$, this
situation may in this sense be qualified of ergodic. For $\delta>2$, the dynamics is governed at large times by a kinematical
term involving the renormalized mass and $\langle v(t_i)\rangle$ is never
forgotten$\protect{\cite{Schramm,Grabert,Aslangul2,Weiss}}$. 

In the following, we limit ourselves to the ergodic case $0<\delta<2$. Then the velocity equilibrates at
large times and does not age. This property, already obtained in the ohmic case$\protect{\cite{Pottier}}$, thus generalizes
to non-ohmic models with $0<\delta<2$. The two-time velocity correlation function depends
only on the time difference and can be computed {\sl via\/} Fourier analysis and the Wiener-Khintchine theorem:
%%%%%%%%%%%%%%%%%%%%%%%
\begin{equation}
C_{vv}(t)={1\over m}\int_{-\infty}^\infty{d\omega\over 2\pi}\,{\Re{\rm
e}\,\tilde K(\omega)\over|\tilde K(\omega)-i\omega|^2}\hbar\omega\coth{\beta\hbar\omega\over 2}e^{-i\omega t}.
\end{equation}
%%%%%%%%%%%%%%%%%%%%%%%
With the modelization (\ref{modelization}) for $K(\omega)$ and a lorentzian cut-off function
$f_c={\omega_c^2/(\omega_c^2+\omega^2)}$, one has:
%%%%%%%%%%%%%%%%%%%%%%%
\begin{eqnarray}
\nonumber &\Im{\rm m}\,\tilde K(\omega)=\omega
{\bigl({|\omega|\over\gamma}\bigr)}^{\delta-2}&f_c\bigl({\omega\over\omega_c}\bigr)\times\\
&&\Bigl[\cot{\delta\pi\over
2}+{\bigl({\omega_c\over|\omega|}\bigr)}^{\delta-2}{1\over\sin{\delta\pi\over 2}}\Bigr].
\end{eqnarray}
%%%%%%%%%%%%%%%%%%%%%%%

Then, setting
%%%%%%%%%%%%%%%%%%%%%%%
\begin{equation}
C_{vv}(\omega)={1\over m}{\Re{\rm
e}\,\tilde K(\omega)\over|\tilde K(\omega)-i\omega|^2}\hbar\omega\coth{\beta\hbar\omega\over 2}, 
\end{equation}
%%%%%%%%%%%%%%%%%%%%%%%
one may attempt to define the coordinate spectral density as $C_{xx}(\omega)={C_{vv}(\omega)/\omega^2}$. If convergent,
the integral $\int_{-\infty}^\infty{(d\omega/2\pi)}\,C_{xx}(\omega)$ represents $\langle x^2(t)\rangle$,
a quantity which must be independent of $t$. Checking the small-$\omega$\break behavior of the integrand with the chosen
modelization for the memory kernel, one sees that this is only possible at $T=0$ for $0<\delta<1$. In this case, the particle
is localized and it makes sense to define its position in an absolute way as $x(t)=\int_{-\infty}^t v(t')dt'$. The two-time
position correlation function 
%%%%%%%%%%%%%%%%%%%%%%%
\begin{equation}
C_{xx}(t,t')={1\over 2}\langle\{x(t),x(t')\}_+\rangle, 
\end{equation}
%%%%%%%%%%%%%%%%%%%%%%%
where the symbol $\{.,.\}_+$ stands for the anticommutator, only depends on $t-t'$ and does
not age. In particular one has 
%%%%%%%%%%%%%%%%%%%%%%%
\begin{equation}
\Delta x^2(t)=2\int_{-\infty}^\infty{d\omega\over 2\pi}\,C_{xx}(\omega)(1-\cos\omega t),
\end{equation}
%%%%%%%%%%%%%%%%%%%%%%%
that is, in terms of the integrated velocity correlation function $D(t)=\int_0^t du\,C_{vv}(u)$,
%%%%%%%%%%%%%%%%%%%%%%%
\begin{equation}
\Delta x^2(t)=\Delta x^2(\infty)-2\int_t^\infty D(t')\,dt'.
\end{equation}
%%%%%%%%%%%%%%%%%%%%%%%
The asymptotic behavior at large times of $D(t)$ allows
to characterize the relaxation of the mean square displacement towards $\Delta x^2(\infty)$. 

In other cases, that is at $T=0$ for $1\leq\delta<2$ and at finite $T$ for $0<\delta<2$, the integral
$\int_{-\infty}^\infty{(d\omega/2\pi)}\,C_{xx}(\omega)$ diverges and $\langle x^2(t)\rangle$ and $C_{xx}(t,t')$ are infinite.
The particle diffuses. Since it is then no more possible to define an absolute position, we focus the interest on the
displacement $x(t)-x(t_0)$ ($t\geq t_0$). This quantity does not equilibrate with the bath, even at large times: it ages. The
mean square displacement $\Delta x^2(t)$ diverges at infinite time and $D(t)$ represents the time-dependent diffusion
coefficient (in an extended sense when diffusion is anomalous, that is at $T=0$ for $1<\delta<2$, and at finite $T$ for
$0<\delta<1$ and $1<\delta<2$). 

Before introducing the appropriate FDT violation factor, let us study in details the
behavior of $D(t)$. Since $0<\delta<2$, one can restrict the study to the infinite bath bandwidth limit
$\omega_c\to\infty$ in which case calculations are more simple. One has:
%%%%%%%%%%%%%%%%%%%%%%%
\begin{eqnarray}
\label{coeffdiffusion}
\nonumber &D(t)=&{\hbar\over m\pi}\gamma^{\delta-2}\int_0^\infty d\omega
\coth{\beta\hbar\omega\over 2}\sin\omega t\times\\
&&\bigl[\omega^{\delta-1}+\omega^{3-\delta}(|\omega|^{\delta-2}\cot{\delta\pi\over
2}-\gamma^{\delta-2})^2\bigr]^{-1}.
\end{eqnarray}
%%%%%%%%%%%%%%%%%%%%%%%
At finite $T$, it is interesting to discuss on the same footing the classical counterpart of $D(t)$, namely 
$D^{\rm cl}(t)$ deduced from $D(t)$ by replacing $\coth{(\beta\hbar\omega/2)}$ by ${2/\beta\hbar\omega}$ in 
formula~(\ref{coeffdiffusion}). Several important features of $D(t)$ and $D^{\rm cl}(t)$ can be obtained by contour
integration.
 
At $T=0$, $D(t)$ is found to be the sum of a pole contribution, which exists only for $0<\delta<1$,
given by the oscillating function
%%%%%%%%%%%%%%%%%%%%%%%
\begin{equation}
D(t)_{\rm pole}\sim{\hbar\over m}{1\over 2-\delta}e^{-\Lambda t}\sin\Omega t,
\end{equation}
%%%%%%%%%%%%%%%%%%%%%%%
where $\Omega$ and $\Lambda$ are known functions of $\delta$ and $\gamma$,
and of a cut contribution behaving at large times as a power-law,
%%%%%%%%%%%%%%%%%%%%%%%
\begin{equation}
\label{Dasympt}
D(t)_{\rm cut}\sim{\hbar\over m\pi}(\gamma t)^{\delta-2}\sin^3{\delta\pi\over
2}\Gamma(2-\delta),
\end{equation}
%%%%%%%%%%%%%%%%%%%%%%%
where $\Gamma$ denotes the Euler Gamma function. 

At finite $T$, $D^{\rm cl}(t)$ is also found to be the sum of an
oscillating function, which exists only for $0<\delta<1$, 
%%%%%%%%%%%%%%%%%%%%%%%
\begin{equation}
D^{\rm cl}(t)_{\rm pole}\sim{kT\over m\gamma}{2\over
2-\delta}{\bigl(\sin{\delta\pi\over 2}\bigr)}^{1\over 2-\delta}e^{-\Lambda t}\sin(\Omega t-\phi)
\end{equation}
%%%%%%%%%%%%%%%%%%%%%%%
with $\phi={\pi\delta/2(2-\delta)}$, and of a cut contribution behaving at large times as a power-law,
%%%%%%%%%%%%%%%%%%%%%%%
\begin{equation}
\label{Dclasympt}
D^{\rm cl}(t)_{\rm cut}\sim{kT\over m\gamma}(\gamma t)^{\delta-1}{\sin{\delta\pi\over 2}\over\Gamma(\delta)}.
\end{equation}
%%%%%%%%%%%%%%%%%%%%%%%

\begin{figure}
\vspace{-4mm}
\epsfxsize=\columnwidth
\centerline{\epsffile{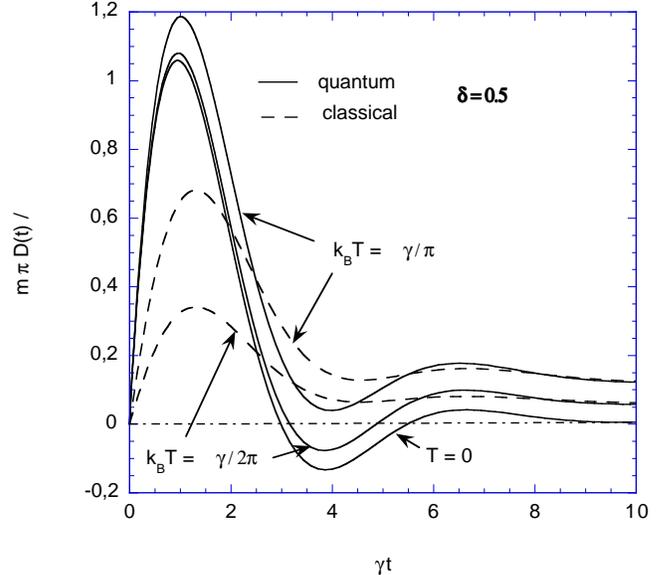}}
\vspace{0mm}
\caption{\label{Figure1}
In full lines, $D(t)$ plotted as a function of $\gamma t$ for
$\delta=0.5$ and for bath temperatures $T=0$, $k_BT={\hbar\gamma/2\pi}$, $k_BT={\hbar\gamma/\pi}$ (at
$T=0$, $D(t)$ is not a diffusion coefficient, but characterizes the relaxation of $\Delta x^2(t)$ towards its limit value). In
dashed lines, the corresponding $D^{\rm cl}(t)$.}
\end{figure}

\begin{figure}
\vspace{-4mm}
\epsfxsize=\columnwidth
\centerline{\epsffile{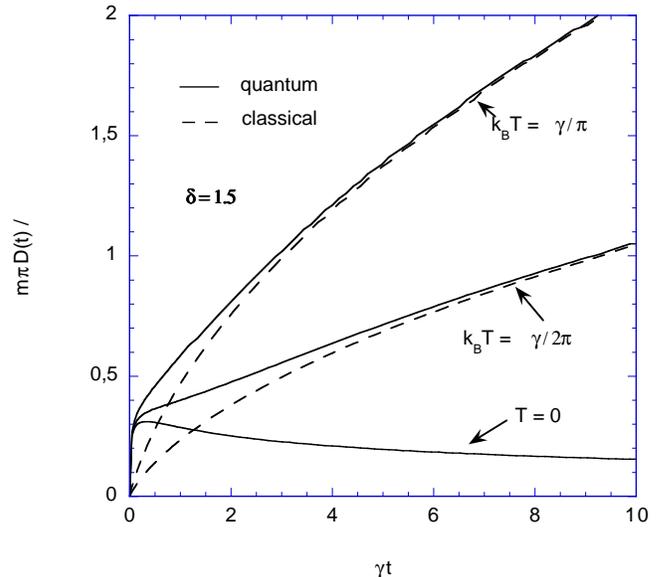}}
\vspace{0mm}
\caption{\label{Figure2}
In full lines, the quantum diffusion coefficient $D(t)$ plotted as a function of $\gamma t$ for
$\delta=1.5$ and for bath temperatures $T=0$, $k_BT={\hbar\gamma/2\pi}$, $k_BT={\hbar\gamma/\pi}$. In dashed lines, the
corresponding 
$D^{\rm cl}(t)$.}
\end{figure}

The behaviors of $D(t)$ and $D^{\rm cl}(t)$ at several different temperatures are illustrated on Fig.~\ref{Figure1} for
$\delta=0.5$ and on Fig.~\ref{Figure2} for $\delta=1.5$. Interestingly enough, for any given~$\delta$, the curves corresponding
to different bath temperatures do not intersect. Actually, it can be shown that, at any fixed time $t$, $D(t)$, like $D^{\rm
cl}(t)$, is a monotonously increasing function of $T$.

For times $t\ll t_{\rm th}$ ($t_{\rm th}={\hbar/2\pi k_BT}$) and for any value of $\delta$, the curves for $D(t)$
at finite $T$ nearly coincide with those at $T=0$, as it should. 

At intermediate times, and for $0<\delta<1$, an oscillation due to the
pole contribution takes place in $D(t)$ (and also in $D^{\rm cl}(t)$ but with a smaller amplitude). For certain values of
$\delta$ and $T$, this oscillation may even result in negative values of $D(t)$ during a finite time interval.
 
At large times
$t\gg\gamma^{-1},t_{\rm th}$ and for any finite $T$, the curves of $D(t)$ and $D^{\rm cl}(t)$ join together: when
\hbox{$0<\delta<1$}, $D(t)$ describes a subdiffusive regime, and, when $\delta>1$, a superdiffusive
one. At $T=0$, for $0<\delta<1$, $D(t)$ describes the relaxation of $\Delta x^2(t)$ towards $\Delta x^2(\infty)$
and, for $1<\delta<2$, a subdiffusive regime. 
%%%%%%%%%%%%%%%%%%%%%%%
%%%%%%%%%%%%%%%%%%%%%%%

%%%%%%%%%%%%%%%%%%%%%%%
%%%%%%%%%%%%%%%%%%%%%%%

These features of $D(t)$, especially the large times ones, are essential for describing the aging properties of the
displacement correlation function, as defined by
%%%%%%%%%%%%%%%%%%%%%%%
\begin{equation}
C_{xx}(t,t';t_0)={1\over 2}\langle\{[x(t)-x(t_0)],[x(t')-x(t_0)]\}_+\rangle.    
\end{equation}
%%%%%%%%%%%%%%%%%%%%%%%

In the classical case, a modified FDT can be written
as$\protect{\cite{Bouchaud,Cugliandolo}}$
%%%%%%%%%%%%%%%%%%%%%%%
\begin{equation}
\chi_{xx}(t,t')=\beta\Theta(t-t')X^{\rm cl}(t,t';t_0){\partial C_{xx}(t,t';t_0)\over\partial t'},
\end{equation}
%%%%%%%%%%%%%%%%%%%%%%%
where $\chi_{xx}(t,t')$ is the displacement response function. For a diffusing particle, the fluctuation-dissipation 
ratio~$X^{\rm cl}(t,t';t_0)$ can be obtained from $D^{\rm cl}(\tau)$ and $D^{\rm
cl}(t_w)$$\protect{\cite{Pottier}}$, where $\tau=t-t'$ denotes the observation time and $t_w=t'-t_0$ the waiting time:
%%%%%%%%%%%%%%%%%%%%%%%
\begin{equation}
\label{X}
X^{\rm cl}(\tau,t_w)={D^{\rm cl}(\tau)\over D^{\rm cl}(\tau)+D^{\rm cl}(t_w)}.
\end{equation}
%%%%%%%%%%%%%%%%%%%%%%%
For any $\tau$ and $t_w$, one can define an effective inverse temperature as $\beta_{\rm eff}^{\rm cl}(\tau,t_w)=\beta X^{\rm
cl}(\tau,t_w)$. Since $X^{\rm cl}$ does not depend on $T$, the bath temperature is rescaled by a factor
${1/X^{\rm cl}}$ larger than 1, due to those fluctuations of the particle displacement which take place during the
waiting time. At large times ($\tau,t_w\gg\gamma^{-1},t_{\rm th}$), one can use in formula (\ref{X}) the asymptotic
expressions of $D^{\rm cl}(\tau)$ and $D^{\rm cl}(t_w)$ as given by Eq. (\ref{Dclasympt}). Eq. (\ref{X}) then displays the
fact that, in a subohmic or superohmic model of exponent $\delta$ ($0<\delta<1$ or $1<\delta<2$), a self-similar aging regime
takes place at large times, as pictured by 
%%%%%%%%%%%%%%%%%%%%%%%
\begin{equation}
\label{selfsimilar}
X^{\rm cl,ag}(\tau,t_w)={1\over 1+{\bigl({t_w\over\tau}\bigr)}^{\delta-1}}.  
\end{equation}
%%%%%%%%%%%%%%%%%%%%%%%
Interestingly enough, $X^{\rm cl,ag}$ and $T_{\rm eff}^{\rm
cl,ag}=(k_B\beta_{\rm eff}^{\rm cl,ag})^{-1}$ are functions of ${t_w/\tau}$, solely parametrized by $\delta$. They do not
depend on the other parameters of the model ({\sl i.e.\/} $\gamma$ or $\omega_c$). For
$\delta=1$, one retrieves the results $X^{\rm cl,ag}={1/2}$ and $T_{\rm eff}^{\rm
cl,ag}=2T$$\protect{\cite{Cugliandolo,Pottier}}$. For any other value of $\delta$, $X^{\rm cl,ag}$ and $T_{\rm eff}^{\rm
cl,ag}$ are algebraic functions of ${t_w/\tau}$. The limits $\tau\to\infty$ and $t_w\to\infty$ do not commute.

In the quantum case, the effective temperature $T_{\rm eff}=(k\beta_{\rm eff})^{-1}$ can be obtained
from the following equation$\protect{\cite{Pottier}}$:
%%%%%%%%%%%%%%%%%%%%%%%
\begin{equation}
\label{Teff}
D_{T_{\rm eff}}(\tau)=D(\tau)+D(t_w).   
\end{equation}
%%%%%%%%%%%%%%%%%%%%%%%
Eq.~(\ref{Teff}) also allows to define $T_{\rm eff}$ at $T=0$ for $1\leq\delta<2$.
Since  $D(t)$ is a monotonously increasing function of $T$, Eq.~(\ref{Teff}) yields for
$T_{\rm eff}(\tau,t_w)$ a uniquely defined value.

\begin{figure}
\vspace{-2mm}
\epsfxsize=\columnwidth
\centerline{\epsffile{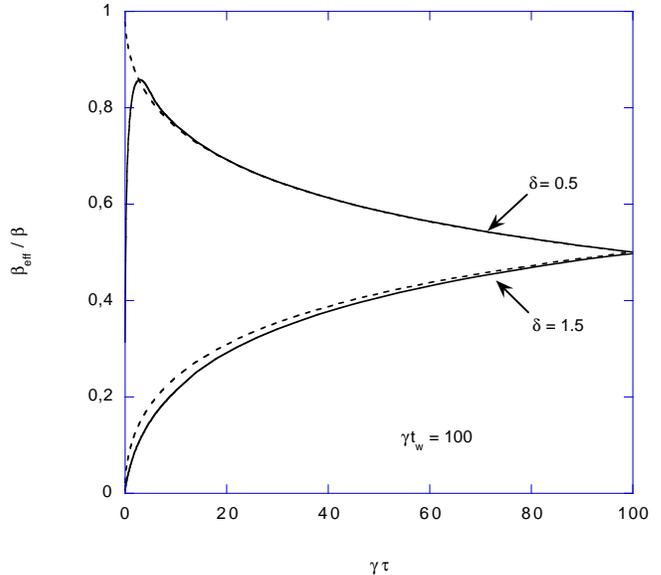}}
\vspace{0mm}
\caption{\label{Figure3}
In full lines, the effective inverse temperature~$\beta_{\rm eff}$, as computed from Eq.~(\ref{Teff}), plotted as a
function of $\gamma\tau$ for a bath temperature $k_BT={\hbar\gamma/2\pi}$ and two different values of $\delta$. In dashed
lines, the corresponding classical effective inverse temperature $\beta_{\rm eff}^{\rm cl,ag}$, as deduced from 
Eq.~(\ref{selfsimilar}).}
\end{figure}

The curves representing $\beta_{\rm eff}(\tau,t_w)$ as a function of~$\tau$ for $\delta=0.5$ and $\delta=1.5$ at a given
finite temperature and for a given $t_w\gg\gamma^{-1},t_{\rm th}$ are plotted on Fig.~\ref{Figure3}. 
Quantum effects do not persist beyond times $\tau\sim t_{\rm th}$. Thus, for times \hbox{$\tau\gg{\gamma}^{-1},t_{\rm th}$,}
$X^{\rm cl,ag}(\tau,t_w)$ as given by formula~(\ref{selfsimilar}) allows for a proper description of aging effects.

In conclusion, we have shown that the two-time dynamics of the coordinate of a quantum dissipative free particle coupled to a
thermal bath (coupling $\sim\omega^\delta$ with $0<\delta<2$) displays extremely rich behaviors. According to the values of
$T$ and $\delta$, one may find either a localized regime in which the position can be defined in an absolute way and does not
age ($T=0,\,0<\delta<1$), or (possibly anomalously) diffusing ones in which it only makes sense to consider the displacement,
which displays aging ($T=0,\,1\leq\delta<2$, or $T$ finite, $0<\delta<2$). The aging regime at finite $T$ is properly
described by the large times expression of the classical fluctuation-dissipation ratio, which for $\delta\neq 1$ is a
self-similar function of
$t_w/\tau$, solely parametrized by $\delta$.

%%%%%%%%%%%%%%%%%%%%%%%
%%%%%%%%%%%%%%%%%%%%%%%

\end{document}